\documentclass{ws-procs961x669}            
\begin{document}
\title{On Einstein's last bid to keep a stationary cosmology}

\author{Salvador Galindo-Uribarri$^{*}$ and Jorge L. Cervantes-Cota$^{**}$}

\address{Departamento de F\'isica, Instituto Nacional de Investigaciones Nucleares,\\
Carretera M\'exico Toluca Km. 36.5, Ocoyoacac, C.P. 52750, Edo. Mex., M\'exico. \\
$^{*}$He recently passed away. \\
$^{**}$E-mail:  jorge.cervantes@inin.gob.mx \\
}

\begin{abstract}
It is commonly known that the steady-state model of the universe was proposed and championed in a series of influential papers around mid-twenty century by Fred Hoyle, Hermann Bondi, and Thomas Gold. In contrast it is little known that, many years before, Albert Einstein briefly explored the same idea; that is of a  ``dynamic steady state" universe. In 1931 during his first visit to Caltech, Einstein tried to develop a model where the universe expanded and where matter was supposed to be continuously created. This latter process was proposed by him to keep the matter density of the universe constant. However, Einstein shortly abandoned the idea. The whole event has already been described and analyzed by C. O'Raifeartaigh and colleagues in 2014. It is the purpose of this brief note to point out what might have prompted Einstein to consider a continuous creation of matter and the prevailing circumstances at that time that drove Einstein's intent.
\end{abstract}


\bodymatter

\section{Introduction}\label{intro}
The first relativistic model of the cosmos is due to Einstein. His model entailed the idea of a universe both isotropic and homogeneous on the largest scales (idea known as ``Cosmological Principle"). At that time, it was the accepted view that the universe was stationary.  This was not unreasonable since relative velocities of stars are small. To achieve a stationary universe, Einstein added in 1917 to his original field equations, the so-called cosmological constant ``$\lambda$" to counterbalance the effects of gravity and attain a static universe\cite{Einstein17}: 

\begin{equation}
G_{\mu \nu} - \lambda \, g_{\mu \nu} =- \kappa \left( T_{\mu \nu} -\frac{1}{2} g_{\mu \nu}  T \right) \, .
\end{equation}

Einstein showed that the value of the constant he introduced was proportional to the mean mass density $\rho$ of the universe and inversely proportional to the square of $R$, its radius of curvature,
\begin{equation}
\lambda  = \frac{\kappa \rho}{2} = \frac{1}{R^2}\, .
\end{equation}

The introduction of $\lambda$ in the original field equations was accepted by the early few practitioners of General Relativity as convenient to keep the current standard view at the time, of a static universe.

However, Einstein was uneasy with his forced introduction of the $\lambda$ constant. Several of his papers show his uneasiness. For instance, in a 1919 publication Einstein stated: 

``...the general theory of relativity requires that the universe be spatially finite. [This] requires the introduction of a new universal constant $\lambda$, standing in a fixed relation to the total mass of the universe (or respectively, to the equilibrium density of matter). This is gravely detrimental to the formal beauty of the theory" \cite{Einstein19} .  

It was not until 1931 that Einstein at long last dropped off $\lambda$ from his field equations. During those days, Einstein was visiting professor in Pasadena California attending an invitation made by Robert Andrews Millikan to spend a short season at Caltech (from late December 1930 to early March 1931). Soon after his arrival to California he discussed Hubble's redshift measurements and its implications with Mount Wilson Observatory astronomers.

Today there is still a justified widespread view that Einstein discarded the cosmological constant immediately after he was satisfied of the validity of Hubble's evidence for a non-static universe. This rendered $\lambda$ in his field equations, redundant. In April of that same year, Einstein submitted for publication, a paper to the Sitzungsberichte der Preussischen Akademie der Wissenschaften where he rejected his cosmological term as superfluous and no longer justified. In his own words, ``theoretically unsatisfactory" 
\cite{Einstein31}.
 
But the view that Einstein dropped off $\lambda$ just before his meetings with the Mount Wilson astronomers (in early Jan 1931) is not strictly exact. In the interval between his arrival to California and his 1931 Sitzungsberichte paper (April 1931), he made a last effort to model a ``dynamic steady state" universe, keeping $\lambda$ in his field equations. This was serendipitously discovered in 2013 by Cormac O'Raifeartaigh in an unpublished Einstein's manuscript kept in the Albert Einstein Archives (AEA) maintained by the Hebrew University of Jerusalem\cite{ORaifeartaigh14a}. The first translation into English of the manuscript, contents and its analysis, has been already covered in 2014 by his discoverer Cormac O'Raifeartaigh and colleagues \cite{ORaifeartaigh14bb}. In addition, the manuscript contents were also commented in a note added in proof by Harry Nussbaumer and later also reviewed by him \cite{Nussbaumer14,Nussbaumer18}. 

\section{A significant Finding}\label{significant}
The discovery by O'Raifeartaigh of the unpublished manuscript showing a model still using the lambda constant, came as a great surprise to everyone.  From the moment Einstein arrived at Caltech it seemed that he had already accepted a non-static universe. Various reports in the local press affirm this. And consequently, people supposed that Einstein had already scrapped the constant from his field equations as unnecessary.    

As a celebrity, during his visit to California, Einstein's activities and sayings were reported daily by the press. In Jan 3 the New York Times (NYT) reported that during an interview given the previous day Einstein stated: ``New observations by Hubble and Humason (astronomers at Mount Wilson) concerning the red shift of light in distant nebulae make the presumptions near that the general structure of the universe is not static" \cite{NYT31_a}. The next following month the NYT (Feb 5) another front-page story, informed that Einstein delivered a lecture the previous day (Feb 4) where he no longer held to the model of ``Stable universe"\cite{NYT31_b}.

On January 15, Einstein attended a welcome dinner in his honor at ``The Athenaeum", a club house for the California Institutes Associates (a group of promoters of southern California scientific and scholarly research). The dinner was attended by around 200 guests among them a selected group of astronomers that had collaborated with their own research in testing relativity\cite{Balch31}. The final dinner speech was delivered by one of them, Walter S Adams, director of Mount Wilson. Adams had verified Einstein's prediction of gravitational redshift\cite{Adams25}. During his speech Adams highlighted the problem of the ``nature and structure of the universe" and he announced that:  

 ``Professor Einstein is now inclined to consider the most promising line of attack on the problem to be based on theories of a non-static universe, the general equations of which have been developed so ably by Dr. Richard Chase Tolman, of the California institute of technology" \cite{Millikan31}. 
 
Despite the multiple examples that can be cited, affirming Einstein's alleged intention to follow a non-static approach, the 2014 discovery by O'Raifeartaigh of the manuscript, revealed that Einstein temporarily followed a different scheme in his unpublished document.

\section{The manuscript }\label{manuscript}
The unpublished manuscript is entitled ``Zum kosmologischen Problem", that is located in the Albert Einstein Archives (AEA) (draft, 1931. Doc [2-112]). It is a signed, four-page handwritten manuscript by Einstein on American paper. Assigned by AEA to January or February 1931.

The cosmological model depicted in the manuscript was not previously detected given that the first words of its title are identical to those of Einstein's 1931 Sitzungsberichte paper (April 1931). For this reason, it was assumed to be a draft of the latter publication.

The manuscript has already been extensively analyzed by O'Raifeartaigh and colleagues \cite{ORaifeartaigh14bb} and in the here cited papers by Nussbaumer, so we shall limit ourselves to giving a succinct description of its contents based on those publications. 

\section{The model in the manuscript} 

In the manuscript Einstein explores a solution to his field equations retaining $\lambda$ that could be compatible with Hubble's observations; to be precise an expanding universe in which the density of matter does not change over time. Einstein starts his analysis by choosing the metric of flat space expanding exponentially (De Sitter metric):
\begin{equation}
ds^2= -e^{\alpha t} (dx_1^2 + dx_2^2 + dx_3^2 )+c^2 dt^2 .
\end{equation}

So, the distance between two points increases over time as $e^{\alpha t/2}$, and he remarks: ``one can thus account for Hubbel's [sic] Doppler effect by giving the masses (thought of as uniformly distributed) constant co-ordinates over time".

In his calculations he finds that $\alpha^2$ (representing the expansion of the universe) is related to its overall density $\rho$ as, 
\begin{equation}
\alpha^2 = \frac{\kappa}{3}  \rho .
\end{equation}

That is, $\rho$ determines the expansion. Since the redshift measurements at the time suggested that the universe expansion is constant, so he concluded that the density must be constant as well.

In the final part of his manuscript, Einstein proposes that the density of matter remains constant by supposing a continuous formation of matter in empty space. He then observes that: ``For the density to remain constant, new particles of matter must be continually formed within that volume from space." 

At the end of the manuscript, he associates the cosmological constant with an energy of space: ``by setting the $\lambda$-term, space itself is not empty of energy". Consequently, the continuous creation of matter becomes associated to the $\lambda$ cosmological constant.

Einstein's ``dynamic steady state" universe simultaneously incorporated the observed expansion of the universe together with a new paradigm, namely of an expanding universe that maintained its isotropy and spatial homogeneity the same as it always has and always will (later dubbed the perfect cosmological principle) by means of the continuous creation of matter. 

The ``dynamic steady state" model was so compelling to set aside. However, Einstein didn't submit his manuscript for publication. O'Raifeartaigh and colleagues detected the possible reason why Einstein gave up publishing it\cite{ORaifeartaigh14bb}. He must have noticed on revision, that his model contained a flaw. Once the error is corrected the model leads to the trivial solution $\rho=0$, that is an empty universe of matter. Therefore, he promptly abandoned this attempt.

We must ask now what it was that attracted Einstein to consider the possibility of a universe where there is a continuous creation of matter. To find a possible answer we shall consider the reception at that time, of an expanding universe as opposed to prevailing static views.

\section{Universe's age and Earth's age tension} 

Towards the beginning of the 1930s the main objection to the interpretation of the Hubble constant as the indicator of the universe expansion was that it's measured value implied a younger universe than that of Earth's accepted age. 

Before 1930's, scientific estimates of the age of the Earth dated from mid-19th century, from Helmholtz-Kelvin gravitational contraction ages\cite{Burchfield90}, passing through those resting on geology (i.e., the amount of salt in the oceans, sediments) and - after a fierce debate between geologist and physicists\cite{Hellman90}  - finally deferred to an age based on early-20th century radioactive dating\cite{Lewis02}. This produced by the 1930's, an estimate of the universe age of around 4 billion years\cite{Badash68}. 

In 1931, at the time Einstein was visiting Caltech, the matter of determining the age of the earth was not entirely resolved to the extent that the National Research Council decided to appoint a committee to investigate to settle the question\cite{Age_earth31}.

On the other hand, the cosmological age of the universe based on the universe expansion, was estimated by taking the reciprocal of the value of the Hubble constant (Hubble's time). Hubble first evaluation of his eponymous constant was of around 500 kilometers per second per megaparsec. This implied a universe's age of about two billion years, which was in a tense contradiction with the estimated age of the Earth, as just mentioned, of about four billion years. As consequence, such mismatch created room for doubt, Einstein himself included. Some critics questioned that the observed nebulae redshifts, were in fact a manifestation of the Doppler's effect. Such was the panorama that reigned in the early 30s.

This incongruity raised two possible explanations: on the one hand Hubble constant value was wrong, or on the other hand, Doppler's shift needed a novel interpretation, as it is said today ``perhaps new physics".

As it is well known, the Hubble constant estimation involved measurement of distance and velocity of a variable star (a Cepheid) belonging to a galaxy in question.  This involved monitoring the apparent brightness of the Cepheid variable to obtain its period. Then, using Shapley's Cepheid calibration, its absolute brightness was established and consequently the distance to the galaxy where the Cepheid was located.  At that time there was little doubt on the correctness of Shapley's calibration. Also, measurement of recession velocity of galaxies even in those days, was straightforward. Spectral recordings had already achieved good accuracy. So, Hubble's constant value was in little doubt. Eventually, some astronomers pointed out that Cepheid's calibration could be slightly inaccurate as the apparent star's luminosity could be diminished by interstellar media and that had to be accounted for. It must be remembered that It was not until mid-1940's that Walter Baade identified two Cepheid types and made a major correction to Shapley's calibration and thus to Hubble's value\cite{Baade44}. Now we know that the problem was on the way off value of Hubble's constant at that time.

On the second possibility, that the incongruity between the ages of the earth and the universe had its origin in an erroneous interpretation of the observed redshift, Fritz Zwicky gave in 1929 an alternative explanation. Zwicky, a resident scholar at Caltech, suggested the concept of ``tired light" \cite{Zwicky29}. This was a hypothetical redshift mechanism where photons lost energy over time through interactions with other particles in their trajectories through a static universe. So, the more distant objects would appear redder than more nearby ones. It is pertinent to mention that the term "tired light" was not used by Zwicky but later coined to refer to this concept by Caltech cosmologist Richard Tolman in the early 1930s in relation to the so called Tolman Surface Brightness Test.

\section{Earlier dynamic stationary state cosmologies. }

Zwicky was not alone. There was also an academic minority that put forward similar ideas to Zwicky's explanation even before his 1929 suggestion. All these ``tired light" propositions had in common the assumption that photons heading earthward, somehow interchanged with the intergalactic medium, part of their energy, thus red shifting their frequencies.

``Tired light" was championed by the few enthusiasts of steady state cosmologies. Its appeal as we shall see resides on the fact that it circumvents the standard interpretation of the redshift as an indicator of galaxy recession and thus safeguards their view of a static universe.

The Nobel laureate Walther Nernst and William Duncan Macmillan, a well-known Chicago professor, were leading advocates of a steady-state universe. What these had in common is that they presuppose the universe as eternal, self-preserving structure in which matter and radiation are constantly been transformed into one another to keep the universe in a stationary balanced state thus avoiding a heat-dead of the universe. A cosmic destiny they both abhorred.

MacMillan's supposes in his own version of ``tired light": ``That there is a leakage of energy from the photon [...] due perhaps to an inherent instability in the photon, or, possibly, [due] to collisions with other photons". He concludes therefore: ``it is evident that the frequency [of the photon] declines with the energy, and the lines of the spectrum are shifted toward the red" \cite{MacMillan32}.

In addition, he considers that the ``evaporated" energy from the photon continues to exist as abundant radiant energy of ``very low frequency".  A kind of primitive version of the modern CMB radiation (but the latter being Black Body radiation). Then he puts forward the possibility that perhaps the ``evaporated" energy ``disappears into the fine structure of space and reappears eventually in the structure of the atom". In other words, he proposes a mechanism of matter creation. 

Another notable hypothesis was that of the Nobel laureate, Walther Nernst, who suggested that some photons are partially absorbed by the luminiferous $\ae$ther  (which he accepts to exist)\cite{Nernst28}. In his 1928 essay ``Phyco-Chemical Considerations in Astrophysics", Nernst outlines the notion of matter creation in his stationary universe:

``I may therefore hold fast to the hypothesis uttered by me that, just as the principle of the stationary condition of the cosmos demands that the radiation of the stars be absorbed by the luminiferous $\ae$ther, so also finally the same thing happens with mass, and that, conversely, strongly active elements are continually being formed from the $\ae$ther, though naturally not in amounts demonstrable to us,..." \cite{Nernst28_a}. 

It is worth noticing that both, MacMillan and Nernst, claim energy transfer occurs by an unknown process yet to be discovered.

The ``tired light" proposal did not vanish into oblivion as alternative explanation to redshift been caused by recessional motion. As late as 1935, Edwin Hubble himself and Richard Tolman (both at Caltech during Einstein's several yearly visits) investigated the possibility that ``tired light" might be an alternative interpretation of redshift. 

They did that comparing the surface brightness of galaxies as a function of their redshift (applying the so called ``Tolman Surface Brightness Test"\cite{Tolman30}). According to Tolman Test, the relationship between surface brightness of a galaxy and its redshift differs in the case of a static universe from that of an expanding one. In a coauthored publication they explored this possibility\cite{Hubble35}. In their joint paper and to simplify their analysis, they employed Einstein's static model of the universe: ... ``combined with the assumption that the photons emitted by a nebula [a galaxy] lose energy on their journey to the observer by some unknown effect, which is linear with distance, and which leads to a decrease in frequency, without appreciable transverse deflection". In short, they employed Einstein's model plus ``tired light". As result of their analysis, they conceded: ``Until further evidence is available, both the present writers wish to express an open mind with respect to the ultimately most satisfactory explanation of the nebular red-shift" ...``They both incline to the opinion, however, that if the red-shift is not due to recessional motion, its explanation will probably involve some quite new physical principles".

\section{On Einstein's arrival at Caltech} 

Up until now, we have seen that in the late 1920s and early 1930s, Hubble's law was commonly interpreted as a demonstration that the universe was truly expanding. However, there was still the problem of reconciling the age of the universe with that of the earth. This inconsistency, in addition to raising a reasonable doubt about the interpretation of the redshift, gave rise to the credibility of the theories of a stationary universe. For some, the appeal of a universe without evolving through time resided in that such universe has no beginning and no end, so it converts the ``age of the universe" as a senseless question issue, consequently eliminating the earth-universe ages paradox.

Attempting to develop a stationary cosmological model was indeed an attractive motivation for Einstein, since it would avoid his well-known aversion to a universe that has a beginning and reinforce his well-known paradigm of a static universe. But adopting such an idea required reliable and robust observational evidence of the continuous creation of matter.

As we will see below, fresh ``evidence" (that turned out to be illusory) of the continuous creation of matter emerged just as Einstein arrived at Caltech. This came from Millikan (Einstein host at Caltech) so it must have seemed reliable to Einstein. We must remember that Millikan received the Nobel prize partly for experimentally verifying Einstein's photoelectric equation.

\section{Birth cries of atoms}

On the 1st of February 1931, during Einstein's stay at Caltech, a paper by Millikan and his collaborator H. Cameron appeared in Physical Review. Its title was ``A more accurate and more extended cosmic-ray ionization -depth curve, and the present evidence for atom -building"\cite{Millikan31_a}. This paper contained what it would seem to be the experimental evidence Einstein needed, that is, observational evidence of continuous matter creation. In our opinion, it was this publication what encouraged Einstein to use the continual creation of matter hypothesis in his unpublished manuscript. So, Einstein made his last bid to keep a stationary cosmology.

Millikan's 1931 paper is the apogee of a series of publications on cosmic rays where he explores their composition and origins. But, before we comment on this publication it is convenient to recall very briefly Millikan's research on cosmic rays.  During the early 20's he and his collaborators made a series of observations on board of balloons, high altitude peaks, and different geographical locations, at various latitudes and in the depths of lakes proving in his 1926 publication that cosmic rays were of extraterrestrial origin\cite{Millikan26}. Millikan believed that his measurements proved that the primary cosmic rays were energetic photons. He also stated that cosmic rays, ``...must arise from nuclear changes of some sort..."  but far more energetic than any radioactive change thus far on record".

In the same 1926 paper he suggested that cosmic rays probably came from among the following three nuclear processes: a) The capture of an electron by the nucleus of a light atom, b) the formation of helium out of hydrogen, or c) some new type of ``nuclear change", such as the condensation of radiation into atoms. 

Regarding third suggestion he made above, we must emphasize that Millikan here raises the creation of matter from the ``condensation of radiation into atoms." this idea was not his, as he himself stated in a note written by him for the journal Science in 1930\cite{Millikan30}. In the note he recalls having in 1915 discussions on the ``running down of the universe" (i.e., when the universe has reached a state of maximum entropy) at the University of Chicago with his then colleague William Duncan Macmillan (of whom we have already commented) where the issue of atom building was discussed. Millikan recalls, 

``In our conversations at Chicago W. D. Macmillan constantly held out for the view that a still further step forward should be taken and that the idea of ``running down of the universe" should be given up by the assumption that atom building went on in space by the condensation of radiation into atoms. He discussed his idea in detail with me in the year 1915, and in July, 1918 he published it in full"\cite{Millikan18}.

In 1928 Millikan and Cameron found that incoming cosmic rays could be grouped into three independent energy bands centered at 26, 110, and 120 (MeV)\cite{Millikan28}. They argued that these bands are produced by the release of photons when eventually a ``sudden union" (i.e., fusion) between atoms occur. To support their assertion, they observed that the center of the bands agreed respectively with 26 MeV (which is just about the mass defect of helium), 110 MeV (which is close to the mass defect of oxygen and nitrogen) and 220 MeV (to that of silicon.) So, their inference was that the three photon bands reaching the earth must be generated by the ``sudden union" of: 4 hydrogen atoms fused to form helium, 14 to form nitrogen, 16 to form oxygen and 28 to form silicon". Secondary electrons, they claimed, were produced in the atmosphere by Compton scattering of gamma rays. Millikan believed that space was filled with a tenuous gas of electrons and protons (the latter he called ``positive electrons"). So, to get around the ``running down of the universe" he assumed that ``These building stones [protons and electrons] are continuously being replenish throughout the heavens by the condensation with the aid of some as yet wholly unknown mechanism of radiant heat into positive and negative electrons".

Millikan supposed discovery of the continuous formation of matter went beyond the scientific sphere provoking the attention of among the scientific and the lay publics. In a statement made by Millikan to the press, cosmic rays were ``birth cries of atoms, a Millikan phrase that achieved a good deal of currency"\cite{Kevles79}.  

Regarding Millikan's 1931 publication the one which appeared on print just at the time Einstein was hosted by Millikan, this was basically a continuation of his 1928 paper including revised energy calculations that further bolstered his tenacious ideas. A short time later Arthur Holly Compton showed that not all cosmic rays were photons, but at least a large part of them consists of charged particles. This led to a debate between the two Nobel laureates. The debate went on for some time with Compton at the winning side but that's another story. 

\section{Final comments}

Einstein before long abandoned the idea of keeping the $\lambda$ constant in his field equations. In his next publication, the 1931 Sitzungsberichte paper, he does not make use of the constant anymore\cite{Einstein31}.  At the end of the paper Einstein adds some remarks about the age of the universe problem, which was quite severe without the use of the $\lambda$ constant. Today we know that the problem was on the way off value of Hubble's constant at that time. Einstein signals two possible errors that may be occurring in the initial approach to the problem. First, he insinuates that the matter distribution might be inhomogeneous and that a homogeneity approximation may be illusional. Then he adds that in astronomy one should be cautious with large extrapolations in time. His first comment is surprising as it seems to indicate some reservations on the cosmological principle.

The next following winter (1931-1932), Einstein was back in Caltech for a second stay where he met De Sitter. Together they formulated a model today known as the Einstein-de Sitter universe, assuming a flat space with no cosmological constant and with an expansion velocity asymptotically approaching zero in the infinite future \cite{Einstein_deSitter32}. This became the standard model up to the mid-1990s. This was the Einstein's last intent to produce a cosmological model. 

\section*{Acknowledgments}
JLCC acknowledges support by CONACyT project 283151.

\end{document}